%% file: neurips_2026.tex
\documentclass{article}

\PassOptionsToPackage{numbers, compress}{natbib}
\usepackage[preprint]{neurips_2026}

\usepackage[utf8]{inputenc} 
\usepackage[T1]{fontenc}    
\usepackage{hyperref}       
\usepackage{url}            
\usepackage{booktabs}       
\usepackage{amsfonts}       
\usepackage{nicefrac}       
\usepackage{microtype}      
\usepackage{xcolor}         
\usepackage{caption}
  \usepackage{graphicx}
  
\usepackage{multirow} 
\usepackage{marvosym}
\usepackage{enumitem}
\usepackage{subcaption}
\usepackage{amsmath} 
\title{WavCube: Unifying Speech Representation for Understanding and Generation via Semantic-Acoustic Joint Modeling}

\author{%
  \textbf{Guanrou Yang$^{1,2}$, Tian Tan$^1$, Qian Chen$^4$, Zhikang Niu$^{1,2}$, Yakun Song$^{1,2}$,} \\
  \textbf{Ziyang Ma$^{1,2}$, Yushen Chen$^{1,2}$, Zeyu Xie$^5$, Tianrui Wang$^3$, Yifan Yang$^1$,} \\
  \textbf{Wenxi Chen$^{1,2}$, Qi Chen$^{1,2}$, Wenrui Liu$^6$, Shan Yang$^3$, Xie Chen$^{1,2}$\thanks{Corresponding author.}} \\
  \\
  $^1$Shanghai Jiao Tong University \quad $^2$Shanghai Innovation Institute \quad $^3$Tencent \\
    $^4$Independent Researcher  \quad $^5$Peking University \quad $^6$Zhejiang University \\
  \texttt{\{yangguanrou,chenxie95\}@sjtu.edu.cn}
}

\begin{document}

\maketitle

\begin{abstract}
Integrating speech understanding and generation is a pivotal step toward building unified speech models. 
However, the different representations required for these two tasks currently pose significant compatibility challenges.
Typically, semantics-oriented features are learned from self-supervised learning (SSL), and acoustic-oriented features from reconstruction. 
Such fragmented representations hinder the realization of truly unified speech systems. 
We present \textbf{WavCube}, a compact continuous latent derived from an SSL speech encoder that simultaneously supports speech understanding, reconstruction, and generation.
WavCube employs a two-stage training scheme.
Stage~1 trains a semantic bottleneck to filter off-manifold redundancy that makes raw SSL features intractable for diffusion.
Stage~2 injects fine-grained acoustic details via end-to-end reconstruction, while a semantic anchoring loss ensures the representation remains grounded within its original semantic manifold.
Comprehensive experiments show that WavCube closely approaches WavLM performance on SUPERB despite an $8\times$ dimensional compression, attains reconstruction quality on par with existing acoustic representations, delivers state-of-the-art zero-shot TTS performance with markedly faster training convergence, and excels in speech enhancement, separation, and voice conversion tasks on the SUPERB-SG benchmark. 
Systematic ablations reveal that WavCube's two-stage recipe resolves two intrinsic flaws of SSL features for generative modeling, paving the way for future unified speech systems. 
Codes and checkpoints are available at \url{https://github.com/yanghaha0908/WavCube}. 
\end{abstract}


\section{Introduction}

Speech processing has achieved remarkable success on a wide spectrum of tasks, from recognition and understanding to generation and speaker modeling. Yet these capabilities are predominantly realized through specialized architectures, each independently optimized for a narrow task family on an independently chosen representation.
In contrast, the vision communities have rapidly converged toward \emph{unified multimodal models} that integrate understanding and generation within a single framework~\citep{zhang2026openvision,fan2025unified,deng2025emerging-bagel,liao2025mogao}. Such unification brings compelling benefits: comprehension and creation mutually reinforce each other---stronger understanding guides higher-quality generation while generative feedback loops facilitate reasoning~\citep{yang2025survey}; a shared representation eliminates the architectural redundancy of separate encoders and resolves the format incompatibility, such as spatial and temporal resolution, and channel dimension across pipelines~\citep{liu2025tuna}; and unified latents further unlock emergent capabilities such as in-context cross-modal interaction and latent-space test-time scaling, where models can reason directly over the generative latent without round-tripping through the pixel decoder~\citep{tong2026scaling,zhao2025unified}. Speech, however, still lags behind this unification trend, largely because understanding and generation have long relied on fundamentally different continuous representations.

At the core of this challenge lies the question of \emph{representation}. On the one hand, Self-supervised learning has reshaped the landscape of speech understanding: wav2vec~2.0~\citep{baevski2020wav2vec}, HuBERT~\citep{hsu2021hubert}, and WavLM~\citep{chen2022wavlm} learn hierarchical features from unlabeled audio that generalize remarkably well across content, speaker, semantic, and paralinguistic tasks~\citep{yang2021superb,mohamed2022self}. These SSL encoders have become the \emph{de facto} substrate for modern speech understanding. 
On the other hand, speech generation predominantly operates on reconstruction-oriented continuous latents such as Mel-spectrograms and VAE-based speech latents~\citep{siuzdak2023vocos,evans2025stable}. While these acoustic representations faithfully preserve fine-grained spectral detail, they inherently encode low-level acoustic variation rather than semantic structure, forcing generative models to learn content, speaker, and prosody from scratch. Worse, this acoustic latent is known to suffer from a \emph{reconstruction-generation dilemma}: enlarging the channel dimension improves reconstruction quality yet simultaneously degrades generative performance, since higher-dimensional unconstrained latents are fundamentally harder for diffusion models to learn~\citep{niu2025semantic, yao2025reconstruction,heek2026unified}. This representational dichotomy where understanding models exploit abstract semantic topologies while generative models are anchored to entangled acoustic details, erects a persistent architectural divide and reinforces the cumbersome dual-tower design that unified multimodal modeling seeks to eliminate~\citep{yan2025ming,tong2026scaling}.

Interestingly, the vision community has recently witnessed a paradigm shift toward \emph{representation-centric generative modeling}. A rapidly growing line of work shows that features from pretrained visual foundation models such as DINOv2~\citep{oquab2023dinov2} and SigLIP~\citep{zhai2023sigmoid} can replace VAE-derived latents~\citep{esser2024scaling} as the substrate for diffusion-based image synthesis. Two recurring observations emerge. First, semantically structured latents are demonstrably more \emph{diffusable}---they accelerate diffusion convergence, enable few-step or even single-step sampling, and narrow the compute-parameter requirement for equivalent sample quality. Second, a carefully prepared semantic latent can simultaneously support discriminative understanding, faithful reconstruction, and high-fidelity generation within a single representation space, enabling the long-sought unification between understanding and generation ~\citep{zheng2025diffusion,zhang2025both,dong2025repack,gao2025one,gong2026rpiae,chen2025aligning,yao2025towards,ye2025distribution,chang2026dino,du2025vqrae, lai2025toward, hu2025meanflow, fan2025prism, pan2025semantics}. The advantages of semantic-centric paradigm extend far beyond static image synthesis: semantic-space reasoning leads to better long-horizon rollouts in world models and navigation~\citep{zhang2026rae,zhang2025efficient}, semantic-space planning enables extreme compression with action-relevant abstraction~\citep{kim2026planning}, and semantic-aware perceptual losses help pixel-space diffusion approach latent-space diffusion~\citep{ma2026pixelgen}. Together these developments raise a compelling question for speech: \emph{can we construct a single, compact continuous latent that simultaneously supports understanding, reconstruction, and generation?}

Realizing this goal in the speech modality is, however, non-trivial. Through systematic diagnosis (Sec.~\ref{sec:Analysis}), we identify two fundamental obstacles inherent to SSL-derived speech representations. \textbf{(i) The high-dimensional redundancy problem.} Directly feeding the 1024-dim WavLM-Large features into a diffusion transformer produces catastrophic failure---a 338M-parameter DiT trained under this setting yields an unreadable WER of 110\% on zero-shot TTS, and even an aggressively scaled 753M variant still exhibits extremely poor acoustic fidelity. This mirrors a widely observed phenomenon in vision: The massive redundancy of SSL feature spaces exacerbates manifold drift in diffusion models, yielding \emph{off-manifold} latents that degrade decoding fidelity. Scaling up the DiT width to brute-force this ambient space is both computationally intractable and fundamentally suboptimal. 
\textbf{(ii) The reconstruction-fidelity gap.} SSL encoders are trained with discriminative objectives that intentionally discard high-frequency, phase-sensitive acoustic cues indispensable for high-fidelity speech synthesis. Naively decoding SSL features therefore yields perceptibly degraded speech. A unified speech latent must reconcile both tensions simultaneously.

To resolve this fundamental dilemma, we propose \textbf{WavCube}, built on a \emph{compress-then-enrich} two-stage recipe that targets these obstacles in turn. 
To attack the redundancy obstacle, Stage~1 utilizes a symmetric adapter-based auto-encoder to distill frozen WavLM-Large features into a 128-dim bottleneck, acting as a principled information bottleneck that carves a compact, diffusion-friendly subspace out of the highly redundant and noisy ambient feature space. In parallel, an acoustic decoder is warmed up via a reconstruction task on the detached latent, ensuring no interference with semantic distillation. 
To overcome the fidelity obstacle, stage~2 unfreezes the SSL encoder and jointly optimizes the entire pipeline with an end-to-end speech reconstruction objective, explicitly injecting fine-grained acoustic detail into the compact latent. A semantic anchoring regularizer strictly tethers both the fine-tuned encoder features and the auto-encoder output to the frozen SSL reference, preventing acoustic enrichment from eroding the well-structured semantic manifold.
WavCube establishes a unified continuous representation where semantic discriminability, acoustic fidelity, and diffusion tractability no longer trade off against one another but coexist as synergetic properties. Our contributions are summarized as follows:
  \begin{itemize}[leftmargin=1.2em, itemsep=2pt, topsep=1pt, parsep=0pt]
    \item We introduce \textbf{WavCube}, a compact continuous representation that unifies speech understanding, reconstruction, and generation within a single space. By infusing fine-grained acoustic details into a distilled SSL semantic manifold, it effectively harmonizes high-level semantic structures with low-level acoustic textures, bridging the long-standing representational gap in speech modeling.
    \item We propose a compress-then-enrich learning recipe designed to resolve the high-dimensional redundancy and acoustic deficit inherent in SSL features, providing a systematic and extensible methodology for transforming discriminative features into unified representations.
    \item Extensive evaluations show WavCube approaches SSL upperbound on SUPERB despite $8\times$ dimensional compression, and matches acoustic representations in reconstruction fidelity. Furthermore, it achieves state-of-the-art zero-shot TTS performance with accelerated training convergence, and consistently outperforms acoustic baselines across SUPERB-SG generation tasks.
\end{itemize}


\section{Related Work}


\subsection{Unified Speech Representations: Prior Efforts and Limitations}

A small but fast-growing line of work targets unified representation that supports both speech understanding and generation. Semantic-VAE~\citep{niu2025semantic} augments VAE with a semantic-alignment regularizer towards pre-trained SSL features. While this successfully mitigates the reconstruction-generation dilemma, the latents are still fundamentally dominated by the reconstruction objective, which limits their capacity for deeper semantic understanding and unified representation modeling.
To address the critical deficiency of Semantic-VAE in speech understanding tasks, JMAS-VAE~\citep{cheng2026distillation} introduces a joint-marginal alignment scheme combined with adaptive loss weighting. This approach explicitly aligns both frame-level features and sequence-level distributions with pre-trained SSL representations. However, achieving a unified representation heavily relies on complex dynamic weighting and carefully calibrated empirical margins to prevent performance collapse. Consequently, the resulting representation is obtained through a fragile, heavily engineered multi-task trade-off rather than an inherently unified structural design.
Dasheng Tokenizer~\citep{dinkel2026dashengtokenizer} freezes a semantic audio encoder and injects acoustic information through a lightweight linear projection, neatly inverting the usual semantic-into-acoustic distillation recipe. Its latent, however, inherits the full encoder dimensionality, which our analysis and several parallel studies identify as intrinsically hostile to diffusion modeling. SemanticVocoder~\citep{xie2026semanticvocoder} discards the VAE altogether and runs flow-matching generation directly in the high-dimensional SSL encoder space, rebalancing difficulty between text-to-latent and latent-to-waveform; yet it shares Dasheng's high-dimensionality burden on the generator.
Ming-UniAudio~\citep{yan2025ming} builds a VAE-based continuous tokenizer with multi-stage LLM-guided semantic distillation; however, it does not achieve a genuinely unified shared latent space for understanding and generation. Specifically, its low-dimensional acoustic representation must pass through an additional semantic module to be transformed into the high-dimensional feature required for understanding tasks. Furthermore, optimizing this decoupled architecture requires a cumbersome three-stage training pipeline comprising acoustic reconstruction, semantic feature distillation, and joint optimization. 


\subsection{Semantic Representations Benefit Generative Modeling}
Parallel efforts in visual representation learning corroborate and inform our speech-domain approach. A first thread establishes that SSL derived latents are fundamentally more diffusion-friendly than traditional reconstruction-trained VAEs. Early explorations approach this by treating pretrained visual models as external supervisors, such as REPA~\citep{yu2024representation}, REPA-E~\citep{leng2025repa}, and VA-VAE~\citep{yao2025reconstruction}.
Pushing this concept further, recent works like RAE~\citep{zheng2025diffusion} and SVG~\citep{shi2025latent} bypass the traditional VAE entirely, proving that strong generative models can be trained directly within the uncompressed, frozen feature spaces of foundation models like DINOv2 and SigLIP, and showing that semantic latents can themselves serve as competitive generative targets.
Complementary to this high-dimensional route, a parallel thread adapts raw SSL features into more generation-friendly latents through a learnable bottleneck and/or reconstruction-driven encoder fine-tuning, such as PS-VAE~\citep{zhang2025both}, RPiAE~\citep{gong2026rpiae}, RePack~\citep{dong2025repack}, FAE~\citep{gao2025one}, Align-Tok~\citep{chen2025aligning}, and DINO-SAE~\citep{chang2026dino}. These methods optimize a low-dimensional bottleneck, optionally coupled with a reference-anchored pixel reconstruction objective. 
Beyond standalone generation, these shared semantic spaces directly enable native unified multimodal models. Models such as OpenVision3~\citep{zhang2026openvision}, Tuna~\citep{liu2025tuna}, and VQRAE~\citep{du2025vqrae} adopt unified visual features to resolve representation format mismatches, unlocking native joint modeling.
Furthermore, semantic-driven paradigms have successfully extended to end-to-end pixel generation, video modeling, and world models.
PixelGen~\citep{ma2026pixelgen} revitalizes end-to-end pixel diffusion by supervising the denoiser with DINOv2-based perceptual losses, steering optimization from the noisy full-image manifold onto a compact perceptual manifold.
For videos, SemanticGen~\citep{bai2025semanticgen} casts video synthesis as a two-stage process that first plans the global layout in a compressed semantic space and then refines high-frequency details in the VAE latent space, yielding faster convergence and better scaling to long videos. DeRA~\citep{guo2025dera} designs a 1D video tokenizer that factorizes video encoding into appearance and motion streams and aligns each stream with a dedicated pretrained vision foundation model.
In world modeling, RAE-NWM~\citep{zhang2026rae} models navigation dynamics directly in dense DINOv2 features, whose superior linear predictability mitigates the structural collapse of compressed latents; ReL-NWM~\citep{zhang2025efficient} runs end-to-end image-goal navigation entirely in DINOv3 space to eliminate costly pixel reconstruction. Planning-in-8-Tokens~\citep{kim2026planning} aggressively resamples frozen foundation features into a highly compact latent that retains only planning-relevant semantics.

\begin{figure}[htbp]
    \centering
    \includegraphics[width=\linewidth]{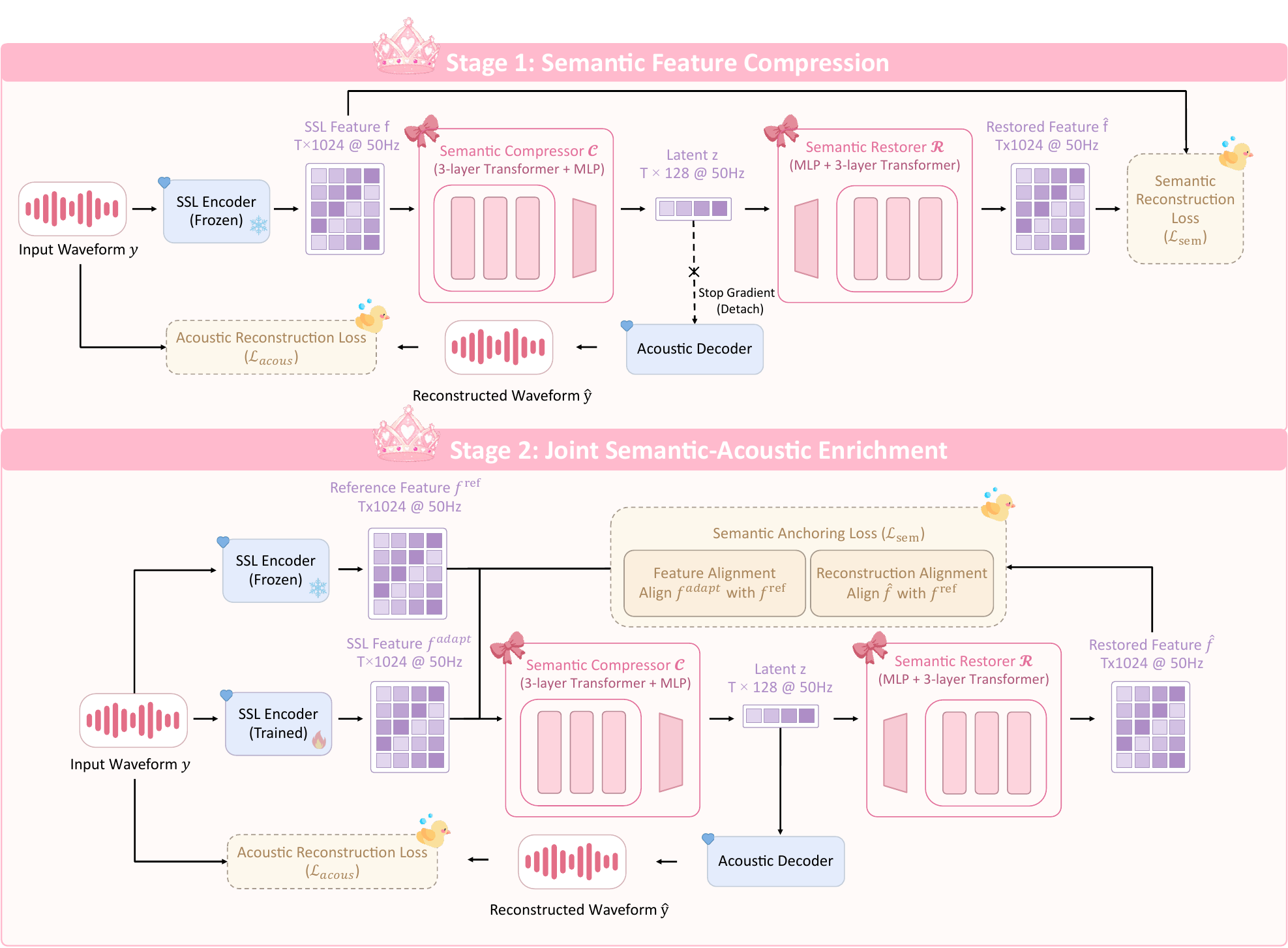}
    \caption{The overall architecture of the WavCube representation. The model is optimized via a two-stage compress-then-enrich paradigm. Stage 1(Top): Semantic Feature Compression. A symmetric auto-encoder compresses features from a frozen SSL encoder into a compact 128-dim latent bottleneck. Concurrently, an acoustic decoder is warmed up on the detached latent to prevent semantic interference. Stage 2(Bottom): Joint Semantic-Acoustic Enrichment. The SSL encoder is unfrozen and the entire pipeline is optimized end-to-end via an acoustic reconstruction loss. A semantic anchoring regularizer strictly aligns both the fine-tuned encoder features and the restored auto-encoder outputs with the frozen SSL reference, injecting fine-grained acoustic details while preventing drift from the original semantic manifold.}

    \label{fig:WavCube_model}
\end{figure}

\section{Methodology}
We present WavCube, a versatile and compact latent representation that demonstrates highly competitive performance across a diverse range of downstream speech tasks, including understanding, reconstruction, and generation. As illustrated in Figure \ref{fig:WavCube_model}, WavCube is built on top of the frozen self-supervised speech encoder WavLM and learned through a two-stage training recipe.

\subsection{Stage 1: Semantic Feature Compression}

Given a $16\mathrm{kHz}$ speech waveform, we first extract its continuous semantic representation using a pre-trained and frozen WavLM model. To bridge the gap between high-dimensional SSL feature and the requirement of efficient downstream generation, we propose a symmetric adapter-based auto-encoder to learn a compact, low-dimensional latent space.

\textbf{Semantic Compressor.} Let $\mathbf{f} \in \mathbb{R}^{T \times d_s}$ denote the sequence of frozen SSL features, where $d_s = 1024$. The compressor module $\mathcal{C}$ maps these features into a bottleneck latent space:
\begin{equation}
    \mathbf{z} = \mathcal{C}(\mathbf{f})
\end{equation}
Specifically, $\mathcal{C}$ consists of a 3-layer Transformer followed by a MLP projection layer. To facilitate faster convergence, the Transformer layers are initialized from the first three layers of the pre-trained WavLM model. The 2-layer MLP projects the sequence into dimensions $d_z = 128$, employing an intermediate dimension of $576$ with GELU activation. The resulting latent $\mathbf{z} \in \mathbb{R}^{T \times d_z}$ maintains a $50\mathrm{Hz}$ temporal resolution, achieving $8\times$ dimension compression relative to the original SSL features.

\textbf{Semantic Restorer.} To ensure the compressed latent $\mathbf{z}$ preserves semantic information and structural integrity, a symmetric restorer module $\mathcal{R}$ is employed to reconstruct the original SSL features:
\begin{equation}
    \hat{\mathbf{f}} = \mathcal{R}(\mathbf{z})
\end{equation}
The restorer architecture mirrors the compressor, comprising a reciprocal projection head and three Transformer layers to lift the $128$-dimensional latent back to the $1024$-dimensional SSL space. This semantic adapter module is optimized by minimizing the Semantic Reconstruction Loss ($\mathcal{L}_{\mathrm{sem}}$). To capture both the magnitude and the directional alignment of the representations, $\mathcal{L}_{\mathrm{sem}}$ is defined as the combination of a Mean Squared Error (MSE) loss and a cosine distance loss between the frozen WavLM features $\mathbf{f}$ and the restored features $\hat{\mathbf{f}}$:
\begin{equation}
    \mathcal{L}_{\mathrm{sem}} = \frac{1}{T} \sum_{t=1}^{T} \left( \left\| \mathbf{f}_t - \hat{\mathbf{f}}_t \right\|_2^2 + 1 - \frac{\mathbf{f}_t \cdot \hat{\mathbf{f}}_t}{\|\mathbf{f}_t\|_2 \|\hat{\mathbf{f}}_t\|_2} \right)
\end{equation}
By optimizing this objective, the adapter strictly distills essential semantic characteristics into the low-dimensional bottleneck.

\textbf{Acoustic Decoder Initialization.} While the primary focus of Stage 1 is establishing the compact semantic latent space, we concurrently perform a preliminary warm-up of the acoustic decoder as an auxiliary task. We adopt the  Transformer-based audio decoder and vocoder from MiMo-AudioTokenizer~\citep{zhang2025mimo}. Taking the detached, dimensionally-reduced latent $\mathbf{z}_{\text{detach}}$ as input, the decoder first projects it to a $1024$-dimensional hidden space via a 1D convolution. Since the latent already operates at the target $50\mathrm{Hz}$ temporal resolution, we bypass the initial temporal upsampling. The sequence is directly processed by $32$ causal Transformer layers. Subsequently, the hidden states are upsampled and mapped to coarse Mel-spectrogram features, which are finally converted into the $16\mathrm{kHz}$ waveform $\hat{\mathbf{y}}$ by the vocoder.
The acoustic reconstruction loss ($\mathcal{L}_{\mathrm{acous}}$) comprises a Mel-spectrogram reconstruction loss ($\mathcal{L}_{\mathrm{mel}}$), alongside adversarial ($\mathcal{L}_{\mathrm{adv}}$) and feature matching losses ($\mathcal{L}_{\mathrm{fm}}$) derived from multi-period and multi-resolution discriminators following Vocos:
\begin{equation}
    \mathcal{L}_{\mathrm{acous}} = \lambda_{\mathrm{mel}} \mathcal{L}_{\mathrm{mel}} + \lambda_{\mathrm{adv}} \mathcal{L}_{\mathrm{adv}} + \lambda_{\mathrm{fm}} \mathcal{L}_{\mathrm{fm}}
\end{equation}
Crucially, since this acoustic loss is computed on the detached latent $\mathbf{z}_{\text{detach}}$, the gradient of $\mathcal{L}_{\mathrm{acous}}$ only updates the acoustic decoder, warming it up in preparation for Stage 2 while leaving the compressed latent shaped purely by the semantic objective.


\subsection{Stage 2: Joint Semantic-Acoustic Enrichment}
 While Stage 1 effectively constructs a compact semantic latent space, its discriminative SSL backbone inherently discards the high-frequency acoustic details that are dispensable for understanding but essential for high-fidelity generation.
To bridge this gap, Stage 2 explicitly injects acoustic information into the semantic latent space through a speech reconstruction objective, while strictly preserving its semantic integrity.

\textbf{Unfreezing the SSL Encoder.} In this stage, we unfreeze the pre-trained WavLM encoder and permits the gradients from the speech reconstruction loss ($\mathcal{L}_{\mathrm{acous}}$) to propagate seamlessly through the semantic compressor $\mathcal{C}$ and into the WavLM encoder. By optimizing the architecture end-to-end via the speech reconstruction task, we fine-tune both the SSL representations and the compact latent bottleneck to capture the fine-grained acoustic details necessary for high-fidelity speech synthesis.

\textbf{Semantic Anchoring.} To prevent the latent space from degrading into a purely acoustic representation, which would severely compromise downstream speech understanding capabilities, we explicitly anchor the fine-tuning process using the original frozen WavLM. Let $\mathbf{f}^{\mathrm{ref}}$ denote the reference features extracted from the frozen model, and $\mathbf{f}^{\mathrm{adapt}}$ denote the representations produced by the actively fine-tuned encoder. The semantic objective is reformulated to encompass two regularization terms. The first is a feature-level constraint that directly aligns the adapted representations $\mathbf{f}^{\mathrm{adapt}}$ with the frozen reference $\mathbf{f}^{\mathrm{ref}}$ to preserve the core semantic information. The second is a reconstruction-level constraint that aligns the restored features $\hat{\mathbf{f}} = \mathcal{R}(\mathcal{C}(\mathbf{f}^{\mathrm{adapt}}))$ with the same frozen reference, ensuring that the auto-encoder bottleneck respects the original semantic manifold. Both regularization objectives employ the combined MSE and cosine distance metric established in Stage 1.

\textbf{Joint Training Objective.} The overall objective for Stage 2 is a weighted summation of the acoustic reconstruction loss and the semantic regularization losses:
\begin{equation}
    \mathcal{L}_{\mathrm{stage2}} =\mathcal{L}_{\mathrm{acous}}(\mathbf{y}, \hat{\mathbf{y}})  +  \lambda_{\mathrm{sem}} \Big( \mathcal{L}_{\mathrm{sem}}(\mathbf{f}^{adapt}, \mathbf{f}^{ref}) + \mathcal{L}_{\mathrm{sem}}(\hat{\mathbf{f}}, \mathbf{f}^{ref}) \Big)
\end{equation}
By jointly optimizing this objective, the architecture achieves a delicate balance between high-level semantics and low-level acoustics. The actively fine-tuned WavLM encoder and the semantic compressor learn to capture rich acoustic details necessary for vocoder synthesis, while remaining bounded by the semantic manifold of the frozen reference. Ultimately, this yields WavCube—a unified representation $\mathbf{z}$ that simultaneously possesses high-level semantic integrity for robust speech understanding, exceptional compactness for efficient generation, and acoustic completeness for high-fidelity reconstruction.

\section{Experiments}
\label{sec:experiments}
To comprehensively evaluate the multifaceted capabilities of the proposed WavCube representation, we design experiments across three distinct dimensions. First, we assess its acoustic fidelity via speech reconstruction task (Sec. 4.1). Second, we evaluate its semantic discriminability using the SUPERB benchmark for speech understanding (Sec. 4.2). Finally, we validate its generative capability through downstream speech generation tasks (Sec. 4.3).

\subsection{Representation Pre-training and Speech Reconstruction}
\subsubsection{Experimental Setup}


\textbf{Datasets.} To evaluate the robustness and scalability of our proposed method, we conduct representation pre-training at two data scales: a standard setting using the 960-hour LibriSpeech dataset~\citep{panayotov2015librispeech} (yielding WavCube), and a 6,000-hour large-scale setting that combines LibriSpeech with the \texttt{small} and \texttt{medium} subsets of the Libriheavy corpus~\citep{kang2024libriheavy} (yielding WavCube-Pro). For evaluation, we consistently report reconstruction performance on the standard LibriSpeech \texttt{test-clean} set.

\textbf{Training Configurations.} We adopt the last hidden layer of the pre-trained WavLM-Large model as the source semantic feature for adaptation. The learning rate follows a linear warmup from $0$ to a peak of $1 \times 10^{-4}$ over the first $5,000$ steps, followed by cosine annealing to $0$. Following the default Vocos configuration, we maintain a $45:1$ relative weighting ratio of $\lambda_{\mathrm{mel}}$ to the adversarial components ($\lambda_{\mathrm{adv}}$ and $\lambda_{\mathrm{fm}}$).
To stabilize the initial generation, Stage 1 optimizes solely the Mel-spectrogram loss for the first $5,000$ steps before introducing adversarial training. In Stage 2, the adversarial objective is applied from the very first iteration, with absolute loss coefficients set to $\lambda_{\mathrm{mel}} = 4.5$, $\lambda_{\mathrm{adv}} = \lambda_{\mathrm{fm}} = 0.1$, and $\lambda_{\mathrm{sem}} = 1.0$.
Following the MiMo-Audio-Tokenizer architecture, our 317M-parameter Acoustic Decoder consists of a 24-layer AudioDecoder with a hidden dimension of 1024, and a 16-layer TransformerVocos that projects these intermediate features into STFT coefficients, enabling the ISTFT head to reconstruct the final 16kHz waveform using an NFFT and window size of 640 with a hop length of 160.

\textbf{Evaluation Metrics.} We assess reconstruction quality across different dimensions: intelligibility using Short-Time Objective Intelligibility (STOI), content consistency using Word Error Rate (WER) computed by Whisper-large-v3 model~\citep{radford2023robust}, perceptual quality via the neural MOS predictor UTMOS~\citep{utmos}, and speaker identity preservation via the cosine similarity (SIM) of speaker embeddings between the ground-truth and reconstructed speech~\citep{spksim}.

\subsubsection{Experimental Result}
The evaluation results for speech reconstruction on the LibriSpeech test-clean set are comprehensively detailed in Table~\ref{tab:reconstruction_metrics}. As expected, acoustic representations such as the Mel-spectrogram, VAE, and Semantic-VAE, which are trained exclusively or predominantly on speech reconstruction tasks, inherently exhibit robust reconstruction performance. However, despite being derived from a semantically structured speech SSL feature under strict semantic regularization, our proposed WavCube representations achieve highly competitive overall reconstruction performance against these acoustic features. 

\input{table/reconstruction}

\subsection{Speech Understanding}
\label{subsec:understanding}

\subsubsection{Experimental Setup} 
To comprehensively evaluate the understanding capabilities and generalizability of WavCube, we utilize the Speech processing Universal PERformance Benchmark (SUPERB). SUPERB is designed to benchmark the performance across ten diverse discriminative tasks, investigating four distinct aspects of speech: content, speaker, semantics, and paralinguistics. Specifically, it encompasses Phoneme Recognition (PR), Keyword Spotting (KS), Query by Example Spoken Term Detection (QbE), and Automatic Speech Recognition (ASR) to probe linguistic content. Speaker characteristics are evaluated through Speaker Identification (SID), Automatic Speaker Verification (ASV), and Speaker Diarization (SD). Furthermore, Intent Classification (IC) and Slot Filling (SF) assess semantic understanding, while Emotion Recognition (ER) tests paralinguistic properties. Following the standard SUPERB framework, we freeze the extracted representations and train only lightweight, task-specific prediction heads akin to linear probing, ensuring that resulting performance strictly reflects the inherent quality of the representations rather than the capacity of the downstream models.

\subsubsection{Experimental Result}
Table~\ref{tab:superb_results} summarizes the SUPERB evaluation results, where WavCube is compared against traditional acoustic filter banks, acoustic representations VAE and Semantic-VAE, and semantic representation WavLM-Large, which serves as the performance upper bound.

Overall, a clear performance hierarchy emerges across SUPERB understanding tasks. The full-dimensional WavLM-Large, recognized as one of the most powerful semantic representations, naturally establishes the performance upperbound. In stark contrast, standard acoustic baselines Fbank, VAE, and Semantic-VAE struggle significantly across all tasks, underscoring their inherent limitations in capturing high-level semantics. Our WavCube comprehensively outperform these acoustic features and achieve highly competitive results that closely follow WavLM-Large. 
This confirms that WavCube successfully preserves the high-level semantics essential for diverse speech understanding tasks.

Although compressing the 1024-dimensional WavLM-Large features into a 128-dimensional latent space inevitably incurs a minor performance drop due to the information bottleneck, the core semantic integrity is well-preserved. Crucially, introducing the acoustic reconstruction objective during the second training stage does not disrupt this semantic structure, for the WavCube and WavCube-Pro representations exhibit negligible fluctuations compared to the semantic-only WavCube-Stage1 feature. 
Besides, scaling the pre-training data from 960 hours (WavCube) to 6000 hours (WavCube-Pro) yields observable improvements across most evaluation tasks, demonstrating the scalability of our unified representation framework.

\input{table/understanding}
\subsection{Speech Generation: Zero-shot Text-to-Speech}

\subsubsection{Experimental Setup}

\textbf{Datasets.} To comprehensively assess representational scalability across different data regimes, we conduct experiments at two distinct scales. For the small-scale evaluation, we utilize the LibriTTS~\citep{zen2019libritts} dataset and report the generation results at $150\mathrm{k}$ training steps. For the large-scale evaluation, we utilize approximately $95,000$ hours English and Chinese speech from the in-the-wild Emilia dataset~\citep{he2024emilia}, filtered for transcription and language errors following the F5-TTS protocol~\citep{chen2025f5}, and evaluate the models at $250\mathrm{k}$ training steps.

\textbf{Training Configurations.} To evaluate the efficacy of various continuous speech representations in downstream generation tasks, we adopt the classic DiT architecture, following the F5-TTS framework. Specifically, our model structure and hyperparameter settings mirror the official \texttt{F5TTS\_v1\_Base} configuration. The DiT backbone features a hidden dimension of $1024$ and a depth of $22$ layers, yielding a total of $337.2\mathrm{M}$ trainable parameters. The models are optimized using a learning rate of $7.5 \times 10^{-5}$ alongside $20,000$ warm-up updates. 

\textbf{Evaluation Metrics.} Following standard evaluation practice, we adopt the LibriSpeech-PC test-clean subset proposed in F5-TTS, which consists of 1,127 audio clips with durations between 4 and 10 seconds. For objective evaluation, we report the WER and Speaker Similarity (SIM-o), computed using the same protocols as in the reconstruction evaluation (Sec. 4.1.1).

\input{table/generation}

\subsubsection{Experimental Result}
To validate the effectiveness of our representation in downstream generative tasks, we conduct a controlled comparison by employing a unified DiT-based TTS architecture and substituting only the underlying continuous speech representations.
As shown in Table \ref{tab:generation_results}, across both the small-scale LibriTTS and the large-scale Emilia-ZH-EN training data configurations, WavCube consistently and significantly outperforms all evaluated baselines, including vanilla VAE, Semantic-VAE, and Mel-spectrograms, in both WER and speaker similarity. 
Specifically, the base WavCube model yields a WER of 1.86\% and a speaker similarity of 0.678 on the LibriTTS dataset, while WavCube-Pro extends this superiority to the Emilia corpus, achieving WER of 2.20\% and speaker similarity of 0.709.
Notably, WavCube achieves this performance operating in the largest 128-dimensional latent space—a factor that theoretically complicates generative modeling, which demonstrates the inherent robustness and architectural advantage of our representation design.

Besides, we compare our system against prominent zero-shot TTS models on large-scale training data configuration. As summarized in Table \ref{tab:sota_comp}, WavCube-Pro exhibits superior performance by consistently outperforming the established baselines CosyVoice~\citep{du2024cosyvoice}, FireRedTTS\citep{guo2024fireredtts}, E2 TTS~\citep{eskimez2024e2} and F5-TTS~\citep{chen2025f5} in both WER and Speaker similarity. (Note that the Mel-spectrogram baseline in Table \ref{tab:generation_results} represents our reproduced performance of F5-TTS, while Table \ref{tab:sota_comp} lists their officially reported results.) This confirms that WavCube serves as a highly competitive representation, effectively driving the TTS system to achieve top-tier performance among contemporary large-scale models.

As illustrated in Figure \ref{fig:convergence}, the convergence curves reveal distinct training trajectories among different representations, with WavCube achieving the fastest convergence speed and the highest stability. From a broader comparative perspective, the semantic-rich WavCube and Semantic-VAE representations consistently optimize much faster than the purely acoustic Mel-spectrogram and vanilla VAE features. 
This phenomenon yields a crucial insight into the training dynamics of diffusion models: high-level semantic representations are fundamentally easier to learn and exhibit noticeably better diffusion-friendly characteristics. Our method provides a highly efficient target space for diffusion-based generative modeling.

\subsection{Speech Generation: SUPERB-SG Generative Tasks}
\subsubsection{Experimental Setup}
\textbf{SUPERB-SG Generation Benchmark.} 
To comprehensively validate the generative capabilities of WavCube across a wider range of generative tasks, we extend our evaluation beyond the zero-shot TTS task. Specifically, we benchmark on three core generative tasks from the SUPERB-SG suite, namely Speech Enhancement (SE), Speech Separation (SS), and Voice Conversion (VC). Following SUPERB, we keep the upstream representations frozen and train only lightweight, task-specific downstream models. 
This probing strategy evaluates the latent space's ability to retain low-level acoustic details and support complex speech generation across varied scenarios.

\subsubsection{Experimental Result}
\input{table/superb_generation}

As shown in Table~\ref{tab:superb_sg_results}, WavLM-Large achieves the best overall performance and serves as the empirical upper bound across these evaluations. Its strong generative capability stems from a joint pre-training paradigm of masked speech prediction and denoising. By predicting clean pseudo-labels from multi-speaker noisy and overlapped speech, WavLM inherently captures robust acoustic and speaker-related priors necessary for non-ASR downstream tasks. Derived by distilling and fine-tuning this powerful model, WavCube naturally inherits these exceptional generative capabilities, establishing a solid foundation for its strong downstream performance.

Interestingly, we observe distinct performance trends across different task categories. For low-level signal reconstruction tasks like Speech Enhancement (SE) and Speech Separation (SS), classic acoustic representations Fbank naturally perform well. 
WavCube achieves highly competitive results, reaching parity with Fbank and clearly outperforming VAE and Semantic-VAE. Furthermore, WavCube excels in Voice Conversion (VC), a task requiring sophisticated decoupling of linguistic content and speaker identity. Besides ensuring a high speaker similarity, WavCube achieves a significantly lower WER compared to other continuous acoustic representations.

\section{Analysis: The Dilemma of SSL Representations and the Role of WavCube Training Stages}
\label{sec:Analysis}
\input{table/analysis}
To deeply understand the architectural necessity of WavCube, we conduct a comprehensive ablation study comparing the reconstruction and generation abilities of the original WavLM feature against our two-stage WavCube representations. The results, summarized in Table \ref{tab:ablation}, reveal crucial insights into the limitations of current SSL features and how our design systematically overcomes them.

Directly utilizing the high-dimensional WavLM for generative tasks exposes two fundamental flaws of current SSL representations. First, it suffers from a severe loss of low-level acoustic details, as evidenced by its poor reconstruction performance with speaker similarity of merely 0.67 and STOI of 0.85. Second, the 1024-dim latent space is overly complex and laden with redundant noise, making it notoriously difficult for diffusion models to learn effectively. When training a standard 339M-parameter DiT whose hidden dimension matches WavLM, the model completely collapses and fails to synthesize intelligible human speech, yielding an unreadable WER of 110.28\%. Although aggressively scaling the DiT hidden dimension to 1536 enables the model to produce intelligible words and reduces the WER to 3.38\%, the overall speech quality remains exceptionally poor, with the speaker similarity sitting at a mere 0.27. Such a dismal acoustic return on a massive 753.5M-parameter endeavor demonstrates that scaling up parameters to brute-force a redundant, high-dimensional SSL space is computationally intractable and fundamentally suboptimal.

To address the modeling difficulties associated with high-dimensional spaces, we compress the 1024-dim WavLM into a compact 128-dim latent space through semantic reconstruction, denoted as WavCube-stage1. This dimensionality reduction yields a vital phenomenon: while reconstruction quality slightly degrades, the downstream TTS generation actually improves, achieving WER of 2.24\%  and speaker similarity of 0.32 with only a lightweight 336M-parameter model. This confirms that dimensionality reduction effectively filters out high-dimensional redundancy, providing a much more diffusion-friendly latent space. However, the speaker similarity remains entirely inadequate, as it is still fundamentally bottlenecked by the inherent lack of acoustic priors inherited from WavLM.

The final WavCube representation resolves this bottleneck through the second stage of acoustic detail injection. This pivotal step transforms WavCube into a highly unified representation that perfectly retains rich semantic structures, possesses ample low-level acoustic details, and maintains a compact, low-dimensional format highly conducive to diffusion modeling. WavCube achieves a near-perfect reconstruction with STOI of 0.97, UTMOS of 4.04, and exceptionally high-fidelity downstream TTS performance with WER of 1.86\% and speaker similarity of 0.68, yielding a comprehensive representation capable of seamlessly bridging speech understanding, reconstruction, and generation.

\section{Conclusions}
We present WavCube, a compact 128-dim continuous representation derived from an SSL speech encoder, capable of unifying robust understanding, high-fidelity waveform reconstruction, state-of-the-art zero-shot TTS, and diverse SUPERB-SG tasks within a single latent space.
The key to this unification is a diagnosis-driven compress-then-enrich recipe: Stage 1 carves a
diffusion-friendly semantic subspace out of the redundant SSL ambient space via an auto-encoder, and Stage 2 injects fine-grained acoustic detail end-to-end while a semantic anchoring regularizer keeps the latent strictly on the SSL semantic manifold. Extensive experiments across diverse speech understanding, reconstruction, and generation benchmarks consistently
demonstrate that semantic discriminability, acoustic fidelity, and diffusion tractability, traditionally viewed as conflicting properties, can coexist as synergetic attributes of a semantically anchored, fine-tuned SSL latent. 
We hope WavCube serves as a foundation for future unified speech modeling, realizing a paradigm where understanding and generation no longer demand separate representational design. To advance this vision, our next step is to construct natively unified speech systems upon WavCube's shared latent representation.

\clearpage

\bibliographystyle{plainnat}
\bibliography{references}
\clearpage

\appendix

\section{Representation Analysis via t-SNE}

To provide a more intuitive understanding of the learned representations, we visualize the latent spaces of WavCube and other baseline features, following MagiCodec~\citep{song2025magicodec}. Specifically, we extract feature sequences using 10 sound categories from the ESC-50 dataset~\citep{piczak2015esc}, aggregate them via mean pooling along the temporal dimension, and project the high-dimensional latents onto a 2D plane using t-SNE.

As illustrated in Figure \ref{fig:tsne}, low-level acoustic representations such as Mel-spectrograms and Acoustic-VAE  exhibit highly entangled distributions with severe overlap across different audio classes. This observation confirms that conventional acoustic latents inherently lack semantic discriminability, as they are primarily optimized for compression and reconstruction. While Semantic-VAE shows marginal improvement, it still fails to form well-defined semantic clusters.
In stark contrast, WavCube groups intra-class samples into compact, well-separated islands, achieving semantic separability on par with the prominent WavLM.
Overall, these visualization results validate that our semantic-acoustic joint modeling effectively preserves rich, high-level semantic structures.

\section{Ablation on Representation Design}

\input{table/tnse}

\input{table/ablation}

Table \ref{tab:appendix_ablation} details the impact of different representation components, specifically the bottleneck architecture, frame rate, latent dimension, and SSL extraction layer, on zero-shot TTS performance.

\textbf{Bottleneck Architecture.} Among three bottleneck designs, the standard Autoencoder (R1) achieves the most favorable trade-off between intelligibility and speaker fidelity, attaining a WER of 2.09\% and a SIM-o of 0.660.
We assume that the KL-divergence penalty in standard VAE (R2) enforces a strict Gaussian prior that over-smooths the latent space. While this constraint might slightly benefit global acoustic modeling, as evidenced by the marginal gain in SIM-o, it may blur the sharp, discriminative boundaries between distinct phonetic units, leading to degraded intelligibility. 
Moreover, the VAE's performance is notoriously sensitive to the KL-divergence weight, requiring delicate hyperparameter tuning to prevent latent collapse. The AE strips away this complex prior matching, offering a structurally simpler and empirically more robust alternative.
Furthermore, we experiment with $\sigma$-VAE (R3) proposed in LatentLM~\citep{sun2024multimodal} to mitigate variance collapse by fixing the variance $\sigma$ to a pre-defined distribution ($\mathcal{N}(0, C_\sigma)$) rather than learning it. However, directly transplanting it to our framework yields sub-optimal results.
 
\textbf{Frame Rate and Latent Dimension.} Regarding temporal resolution, reducing the frame rate from 50Hz to 25Hz (R4) leads to a noticeable decline in both metrics, primarily attributed to temporal information loss. For the latent dimension, compressing the space to 64 dimensions (R5) yields a slight improvement in WER of 1.98\% but noticeably harms speaker similarity. This trade-off is consistent with the conclusions drawn in Semantic-VAE~\citep{niu2025semantic} that increasing the latent dimension provides the necessary capacity to capture rich acoustic details, improving SIM-o, but introduces redundancy that complicates semantic modeling, thereby worsening the WER.

\textbf{SSL Extraction Layer.} Extracting features from the 23rd SSL layer (R6) produces results comparable to the 24th layer, indicating that within our specific architecture, the semantic and acoustic capacities of these two upper layers are similarly effective.

Ultimately, comprehensively weighing the trade-off between content consistency and speaker fidelity, we select the AE bottleneck, 50Hz frame rate, 128 dimensions, and the 24th SSL layer as our optimal default configuration (R1).




\newpage

\end{document}

%% file: table/reconstruction.tex
\begin{table*}[t]
\centering
\caption{Speech reconstruction performance of different continuous speech representations on LibriSpeech test-clean set. }
\label{tab:reconstruction_metrics}
\begin{tabular}{lcccccc}
\toprule
\textbf{Representation} & \textbf{Training Data (hrs)} & \textbf{STOI $\uparrow$} & \textbf{UTMOS $\uparrow$} & \textbf{SIM $\uparrow$} & \textbf{WER(\%) $\downarrow$} \\
\midrule
Ground Truth & - & 1.00 & 4.09 & 1.00 & 3.64 \\
\midrule
Mel-spectrogram & 585 & 0.98 & 3.63 & 0.93 & 3.86 \\
VAE & 6000 & 0.98 & 4.13 & 0.97 & 4.07 \\
Semantic-VAE & 6000 & 0.98 & 4.13 & 0.97 & 4.07 \\
\midrule
\textbf{WavCube}  & 960 & 0.97 & 4.04 & 0.94 & 4.20 \\
\textbf{WavCube-Pro} & 6000 & 0.97 & 4.00 & 0.95 & 4.12 \\
\bottomrule
\end{tabular}%
\end{table*}


%% file: table/understanding.tex
\begin{table*}[t]
\centering
\caption{Speech understanding performance of different continuous speech representations on the SUPERB benchmark.}
\label{tab:superb_results}
\resizebox{\textwidth}{!}{%
\begin{tabular}{lcccccccccccc}
\toprule
\multirow{2}{*}{\textbf{Representation}} & \multirow{2}{*}{\textbf{Dim.}} & \textbf{PR} & \textbf{KS} & \textbf{IC} & \textbf{SID} & \textbf{ER} & \textbf{ASR} & \textbf{QbE} & \multicolumn{2}{c}{\textbf{SF}} & \textbf{ASV} & \textbf{SD} \\
& & PER $\downarrow$ & Acc $\uparrow$ & Acc $\uparrow$ & Acc $\uparrow$ & Acc $\uparrow$ & WER $\downarrow$ & MTWV $\uparrow$ & F1 $\uparrow$ & CER $\downarrow$ & EER $\downarrow$ & DER $\downarrow$ \\
\midrule
Fbank & 80 & 83.71 & 8.85 & 10.16 & 0.06 & 25.62 & 37.95 & 0.0043 & 64.22 & 59.05 & 10.36 & 15.28 \\
VAE & 64 & 88.53 & 39.94 & 9.94 & 15.94 & 44.70 & 63.12 & 0.0002 & 58.93 & 65.55 & 15.04 & 16.57 \\
Semantic-VAE & 64 & 87.59 & 45.30 & 10.63 & 16.40 & 47.28 & 64.64 & 0.0000 & 50.78 & 72.27 & 14.10 & 15.94 \\
\midrule
WavCube & 128 & \underline{9.91} & \textbf{97.42} & \textbf{90.41} & \textbf{42.36} & \underline{63.41} & \underline{9.36} & \underline{0.0367} & \textbf{87.19} & \textbf{28.80} & \textbf{5.86} & \underline{8.14} \\
WavCube-Pro & 128 & \textbf{9.74} & \underline{97.18} & \underline{88.96} & \underline{40.89} & \textbf{66.27} & \textbf{9.34} & \textbf{0.0391} & \underline{86.95} & \underline{28.86} & \underline{6.02} & \textbf{7.77} \\
\midrule
\textcolor{gray}{WavCube-Stage1} & \textcolor{gray}{128} & \textcolor{gray}{8.68} & \textcolor{gray}{96.73} & \textcolor{gray}{91.58} & \textcolor{gray}{38.20} & \textcolor{gray}{64.15} & \textcolor{gray}{6.91} & \textcolor{gray}{0.0488} & \textcolor{gray}{89.19} & \textcolor{gray}{24.70} & \textcolor{gray}{7.35} & \textcolor{gray}{7.44} \\
\textcolor{gray}{WavLM-Large} & \textcolor{gray}{1024} & \textcolor{gray}{3.23} & \textcolor{gray}{98.12} & \textcolor{gray}{100.00} & \textcolor{gray}{93.78} & \textcolor{gray}{70.05} & \textcolor{gray}{3.70} & \textcolor{gray}{0.0532} & \textcolor{gray}{93.49} & \textcolor{gray}{16.92} & \textcolor{gray}{4.93} & \textcolor{gray}{4.00} \\
\bottomrule
\end{tabular}%
}
\end{table*}

%% file: table/generation.tex
\begin{table}[t] 
\centering

\begin{minipage}[t]{0.49\textwidth}
    \centering
    \caption{Zero-shot TTS performance comparison among different continuous speech representations on the LibriSpeech-PC test-clean set.}
    \label{tab:generation_results}
    \resizebox{\textwidth}{!}{ 
    \begin{tabular}{lcccc}
    \toprule
    \textbf{Representation} & \textbf{Dim.} & \textbf{\# Recon. Data} & \textbf{WER $\downarrow$} & \textbf{SIM-o $\uparrow$} \\
    \midrule
    \multicolumn{5}{c}{\textit{TTS Training Data: LibriTTS}} \\
    \midrule
    VAE             & 64  & 6000h & 2.10 & 0.593 \\
    Semantic-VAE    & 64  & 6000h & 2.25 & 0.626 \\
    Mel-spectrogram & 100 & 585h  & 2.02 & 0.598 \\
    \textbf{WavCube} & 128 & 960h  & \textbf{1.86} & \textbf{0.678} \\
    \midrule
    \multicolumn{5}{c}{\textit{TTS Training Data: Emilia-ZH-EN}} \\
    \midrule
    VAE             & 64  & 6000h & 2.47 & 0.673 \\
    Semantic-VAE    & 64  & 6000h & 2.35 & 0.706 \\
    Mel-spectrogram & 100 & 585h  & 2.29 & 0.628 \\
    \textbf{WavCube-Pro}& 128 & 6000h & \textbf{2.20} & \textbf{0.709} \\
    \bottomrule
    \end{tabular}
    }
\end{minipage}
\hfill 
\begin{minipage}[t]{0.49\textwidth}
    \centering
    \caption{System-level zero-shot TTS performance comparison with representative large-scale baselines. Baseline results are cited from F5-TTS.}
    \label{tab:sota_comp}
    \resizebox{\textwidth}{!}{ 
    \begin{tabular}{lcccc}
    \toprule
    \textbf{Model} & \textbf{\# Params.} & \textbf{\# TTS Data} & \textbf{WER $\downarrow$} & \textbf{SIM-o $\uparrow$} \\
     \midrule
     Ground Truth                & -    & -      & 2.23 & 0.690 \\
    \midrule
    CosyVoice  & 300M & 170k h & 3.59 & 0.660 \\
    FireRedTTS & 580M & 248k h & 2.69 & 0.470 \\
    E2 TTS       & 333M & 95k h  & 2.95 & 0.690 \\
    F5-TTS  & 336M & 95k h  & 2.42 & 0.660 \\
    \midrule
    \textbf{WavCube-Pro}         & 337M & 95k h  & \textbf{2.20} & \textbf{0.709} \\
   
    \bottomrule
    \end{tabular}
    }
    
    \vspace{4mm} 
\end{minipage}

\vspace{6mm} 

\begin{minipage}{\textwidth}
    \centering
    \includegraphics[width=\textwidth]{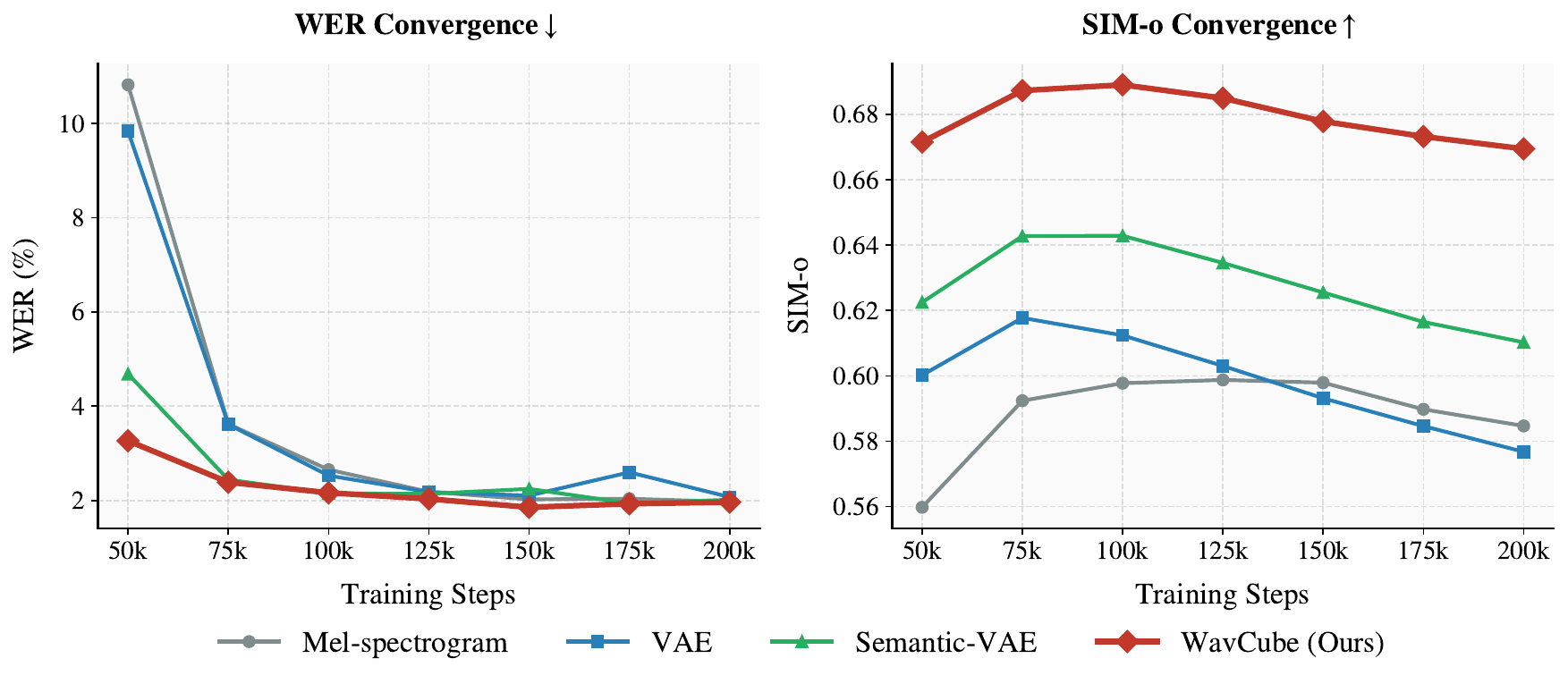} 
    \captionof{figure}{Convergence analysis of Word Error Rate (WER) and Speaker Similarity (SIM-o) during TTS training. WavCube (red) exhibits significantly faster convergence and higher stability compared to other continuous speech representations.}
    \label{fig:convergence}
\end{minipage}
\vspace{-2em}
\end{table}


%% file: table/superb_generation.tex
\begin{table*}[t]
\centering
\caption{Speech generation performance of different continuous speech representations on the SUPERB-SG benchmark.}
\label{tab:superb_sg_results}
\begin{tabular}{lcccccc}
\toprule
\multirow{2}{*}{\textbf{Representation}} & \multicolumn{2}{c}{\textbf{SE}} & \textbf{SS} & \multicolumn{3}{c}{\textbf{VC}} \\
\cmidrule(lr){2-3} \cmidrule(lr){4-4} \cmidrule(lr){5-7}
& PESQ $\uparrow$ & STOI $\uparrow$ & SI-SDRi $\uparrow$ & MCD $\downarrow$ & WER $\downarrow$ & ASV $\uparrow$ \\
\midrule
Fbank & \textbf{2.11} & \textbf{86.2} & \textbf{9.75} & 8.80 & 40.1 & \textbf{72} \\
VAE & 1.89 & 84.8 & 7.76 & 8.77 & 38.6 & 65 \\
Semantic-VAE & 1.90 & 84.9 & 7.37 & 8.90 & 32.6 & 60 \\
\midrule
WavCube & \underline{2.08} & \underline{86.1} & \underline{9.20} & \underline{8.58} & \underline{24.9} & 67 \\
WavCube-Pro & 2.07 & \textbf{86.2} & 9.16 & \textbf{8.43} & \textbf{18.7} & \underline{71} \\

\midrule
\textcolor{gray}{WavCube-stage1} & \textcolor{gray}{1.92} & \textcolor{gray}{84.6} & \textcolor{gray}{5.97} & \textcolor{gray}{7.26} & \textcolor{gray}{11.0} & \textcolor{gray}{100} \\
\textcolor{gray}{WavLM-Large} & \textcolor{gray}{2.18} & \textcolor{gray}{87.1} & \textcolor{gray}{11.23} & \textcolor{gray}{7.65} & \textcolor{gray}{9.8} & \textcolor{gray}{96} \\
\bottomrule
\vspace{-2em}
\end{tabular}
\end{table*}


%% file: table/analysis.tex
\begin{table*}[t]
\centering
\caption{Ablation analysis of representation capabilities across reconstruction and zero-shot TTS tasks. Comparing the original WavLM with WavCube variants demonstrates that high-dimensional SSL features suffer from intractable redundant noise and severe acoustic loss. Our two-stage approach of initial dimensionality reduction followed by acoustic detail injection effectively filters redundant noise and bridges the acoustic gap, yielding a compact, unified representation that masters speech reconstruction and generation.}
\label{tab:ablation}
\resizebox{\textwidth}{!}{
\begin{tabular}{lccccccccc}
\toprule
\multirow{2}{*}{\textbf{Representation}} & \multirow{2}{*}{\textbf{Rep. Dim}} & \multicolumn{4}{c}{\textbf{Reconstruction}} & \multicolumn{4}{c}{\textbf{Zero-shot TTS}} \\
\cmidrule(lr){3-6} \cmidrule(lr){7-10}
 & & \textbf{STOI $\uparrow$} & \textbf{UTMOS $\uparrow$} & \textbf{WER $\downarrow$} & \textbf{SIM $\uparrow$} & \textbf{DiT Dim} & \textbf{\# Params} & \textbf{WER (\%) $\downarrow$} & \textbf{SIM-o $\uparrow$} \\
\midrule
\multirow{2}{*}{WavLM-Large} & \multirow{2}{*}{1024} & \multirow{2}{*}{0.85} & \multirow{2}{*}{3.70} & \multirow{2}{*}{4.09} & \multirow{2}{*}{0.67} 
& 1024 & 338.7M & 110.28& 0.09 \\
& & & & & & 1536 & 753.5M & 3.38 & 0.27 \\
\midrule
WavCube-Stage1 & 128 & 0.81 & 3.10 & 4.40 & 0.54 
& 1024 & 335.9M & 2.24 & 0.32 \\
\textbf{WavCube} & 128 & \textbf{0.97} & \textbf{4.04} & \textbf{4.20} & \textbf{0.94} 
& 1024 & 335.9M & \textbf{1.86} & \textbf{0.68} \\
\bottomrule
\end{tabular}
}
\end{table*}


%% file: table/tnse.tex
\begin{figure*}[t]

      \centering
      \begin{subfigure}[b]{0.33\textwidth}
          \centering
          \includegraphics[width=\linewidth]{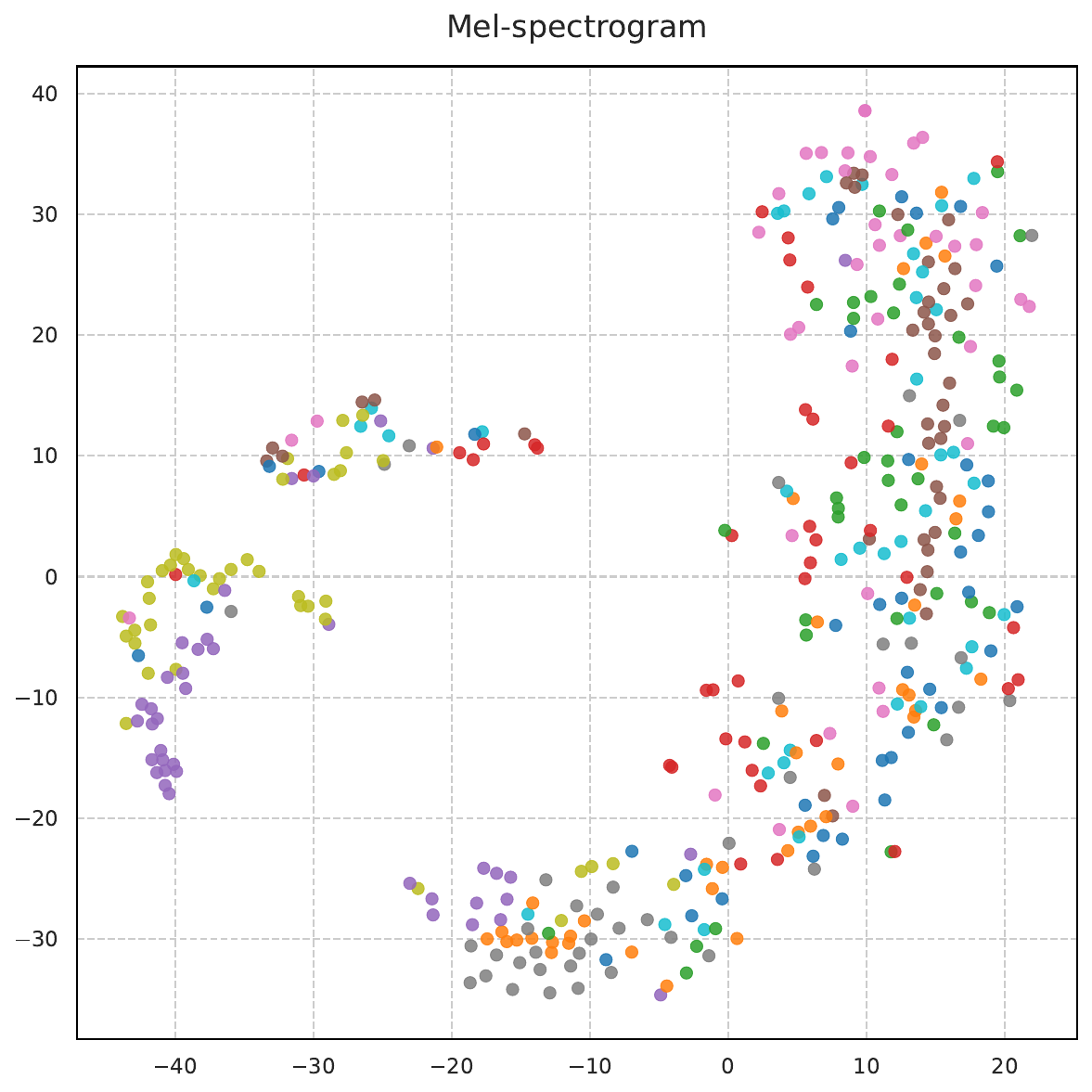}
          \caption*{(a) Mel-spectrogram}
      \end{subfigure}\hfill
    \begin{subfigure}[b]{0.33\textwidth}
          \centering
          \includegraphics[width=\linewidth]{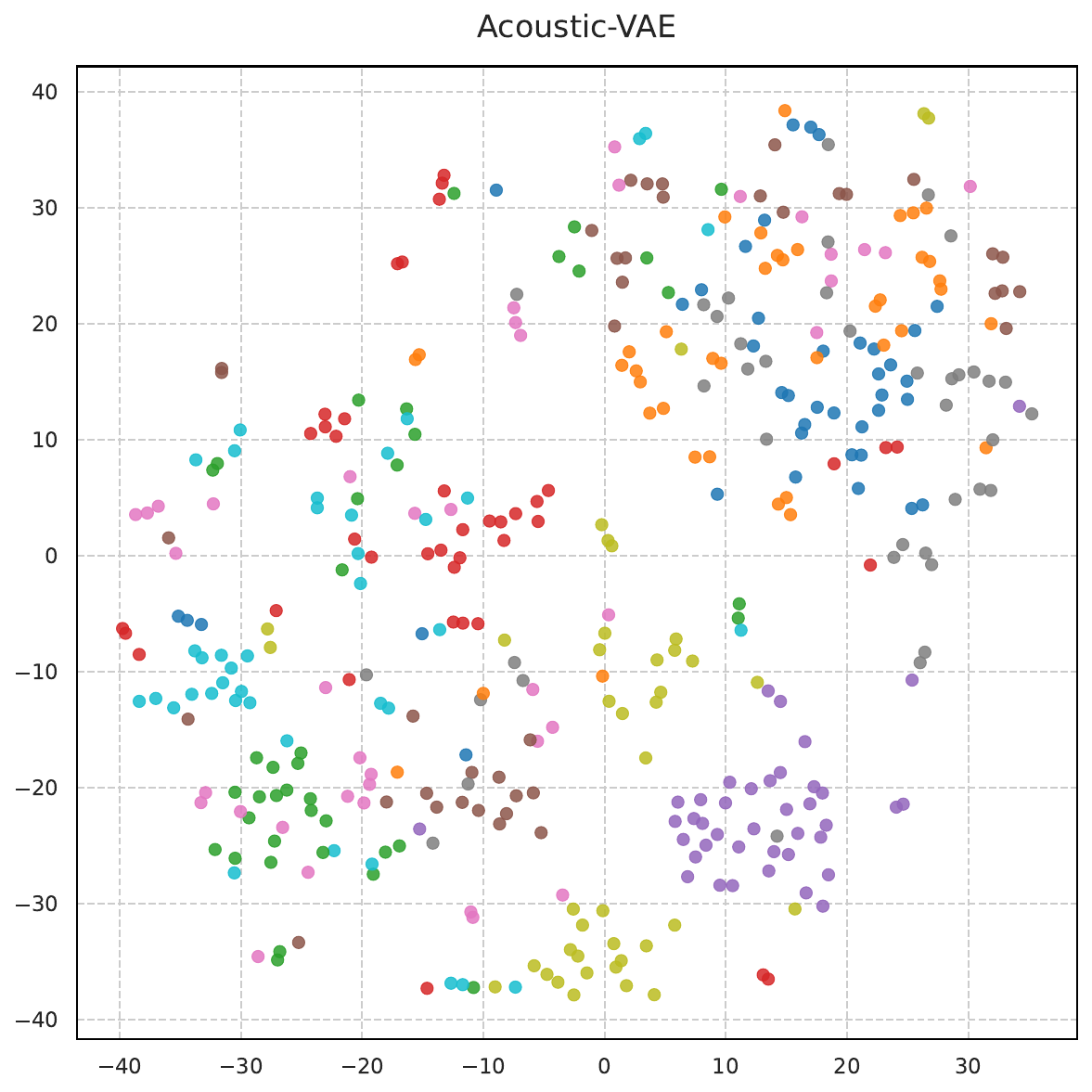}
          \caption*{(b) Acoustic-VAE}
      \end{subfigure}
      \begin{subfigure}[b]{0.33\textwidth}
          \centering
          \includegraphics[width=\linewidth]{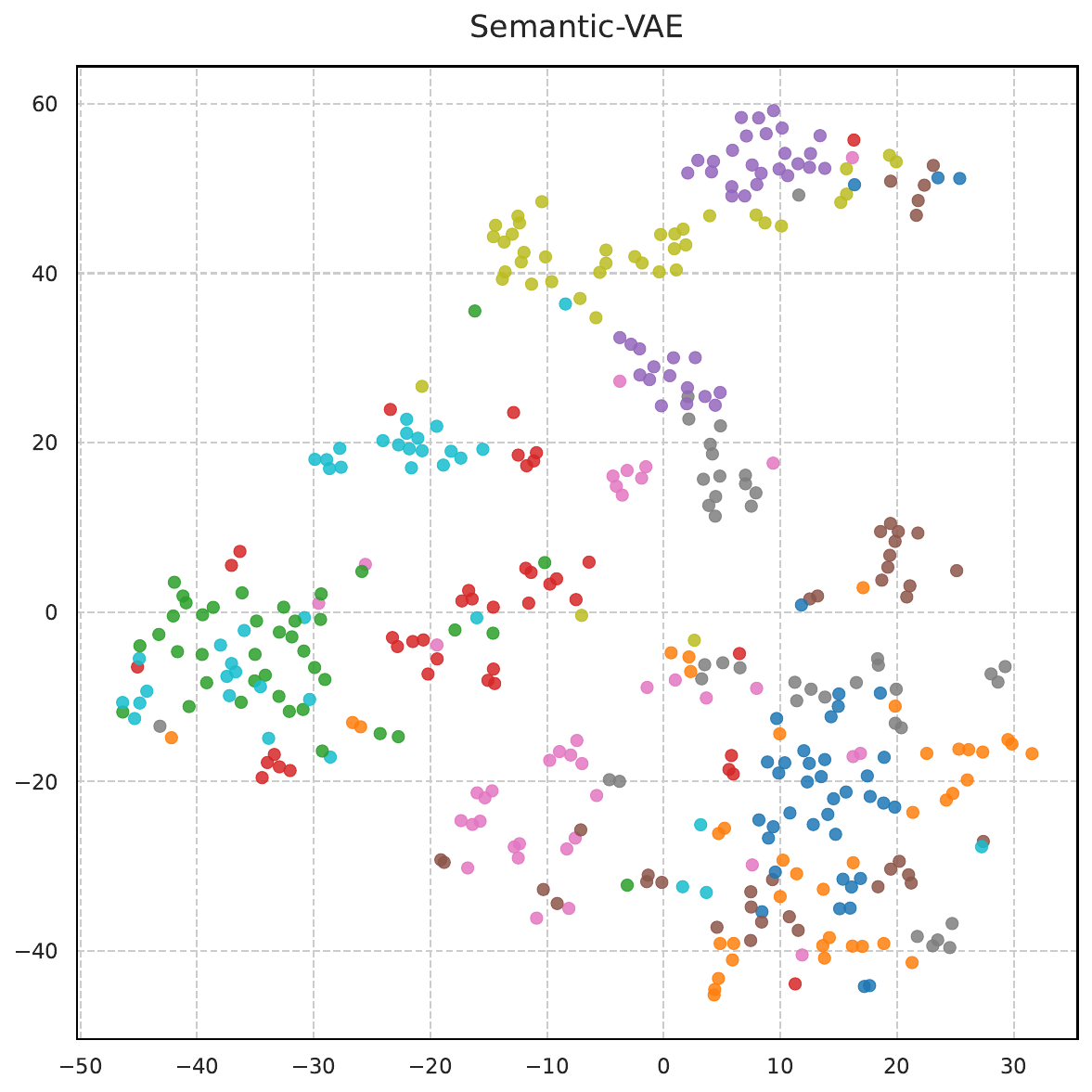}
          \caption*{(c) Semantic-VAE}
      \end{subfigure}\hfill


      \begin{subfigure}[b]{0.33\textwidth}
          \centering
          \includegraphics[width=\linewidth]{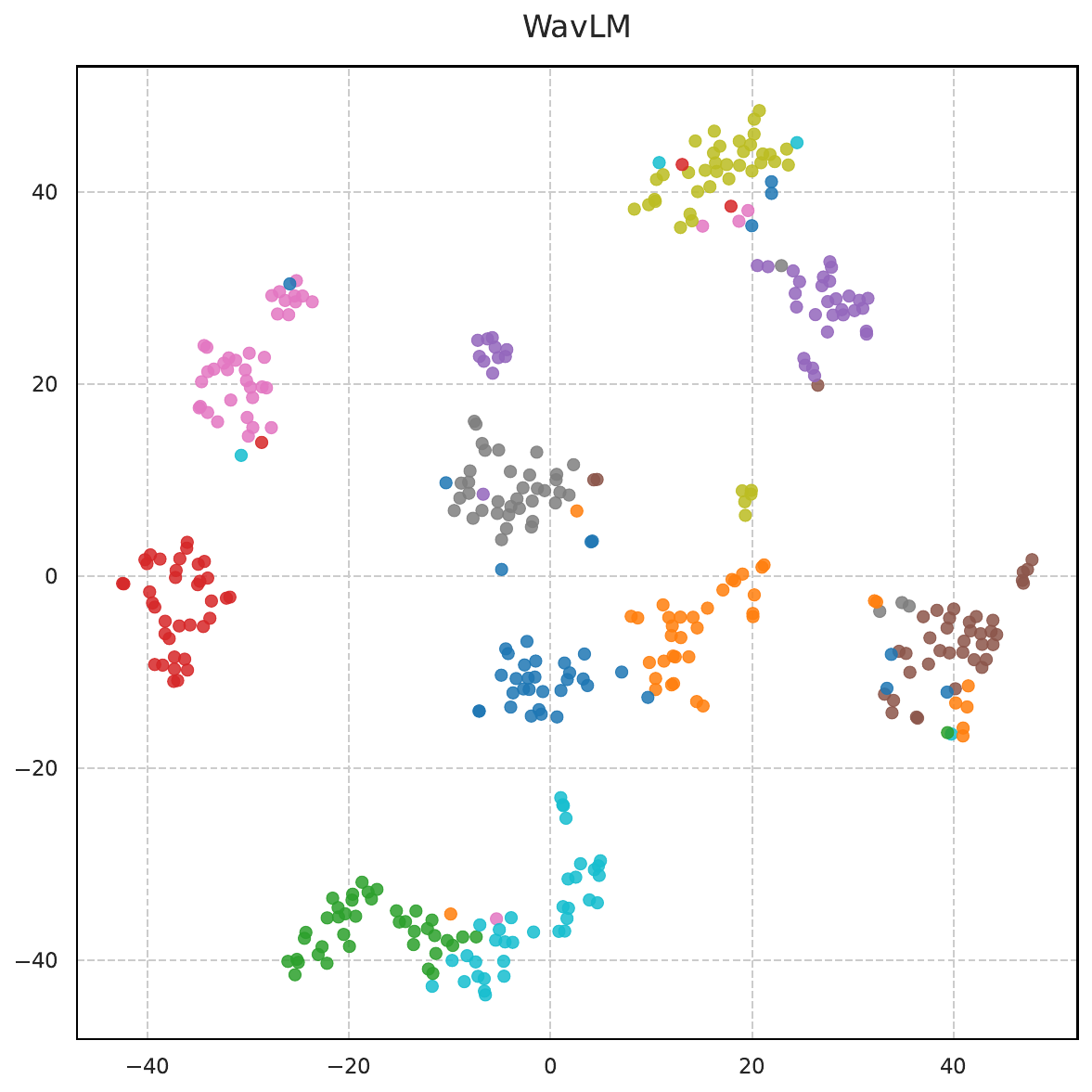}
          \caption*{(d) WavLM}
      \end{subfigure}\hfill
      \begin{subfigure}[b]{0.33\textwidth}
          \centering
          \includegraphics[width=\linewidth]{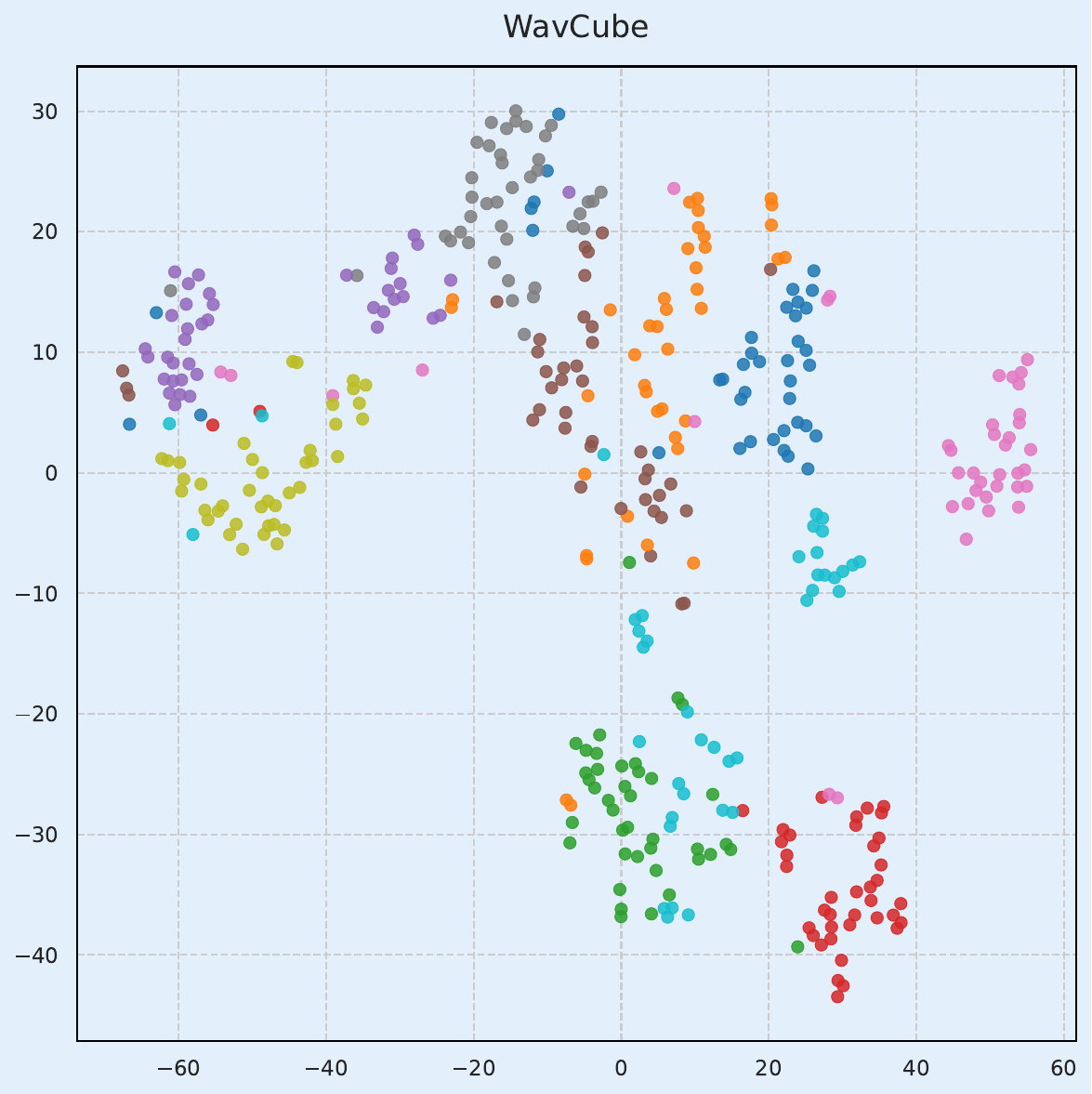}
          \caption*{(e) WavCube (Ours)}
      \end{subfigure}\hfill
      \begin{subfigure}[b]{0.33\textwidth}
          \centering
          \includegraphics[width=0.5\linewidth]{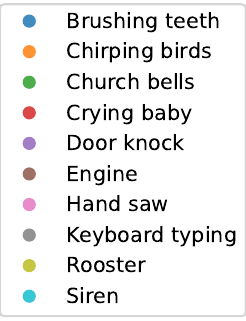}
          \vspace{2em}
      \end{subfigure}
            
\caption{Visualization of different representations on ESC-50 dataset, where 10 representative categories are presented. Each speech feature sequence is aggregated by mean pooling along the temporal axis and projected into 2D space via t-SNE. Compared to Mel-spectrograms and VAE-based baselines, WavCube exhibits a more discriminative clustering structure with enhanced intra-class compactness and clear inter-class margins, achieving on-par semantic separability with WavLM.}
\label{fig:tsne}
  \end{figure*}

%% file: table/ablation.tex
\begin{table}[htbp]
    \centering
\caption{Ablation studies investigating the bottleneck architecture, frame rate, latent dimension, and SSL extraction layer of the WavCube representation. For computational efficiency, we employ a lightweight 19.3M-parameter Acoustic decoder, while all other configurations remain identical to the main experiments. R1 represents the proposed default setting.}
    \label{tab:appendix_ablation}
    \begin{tabular}{ccccccc}
        \toprule
        \textbf{Representation} & \textbf{Bottleneck} & \textbf{Rate} & \textbf{Dim} & \textbf{SSL layer} & \textbf{WER(\%) $\downarrow$} & \textbf{SIM-o $\uparrow$} \\
        \midrule
        \textbf{R1} & AE & 50Hz & 128 & 24 & 2.09 & 0.660 \\
        \midrule
        R2 & \underline{VAE} & \multirow{2}{*}{50Hz} & \multirow{2}{*}{128} & \multirow{2}{*}{24} & 2.36 & \textbf{0.667} \\
        R3 & \underline{$\sigma$-VAE} & & & & 4.49 & 0.658 \\
        \midrule
        R4 & \multirow{2}{*}{AE} & \underline{25Hz} & 128 & \multirow{2}{*}{24} & 2.36 & 0.638 \\
        R5 & & 50Hz & \underline{64} & & 1.98 & 0.581 \\
        \midrule
        R6 & AE & 50Hz & 128 & \underline{23} & \textbf{1.97} & 0.643 \\
        \bottomrule
    \end{tabular}
\end{table}